%% file: main.tex
\documentclass[sigconf,screen]{acmart}

\usepackage{dcolumn}
\usepackage{graphicx}
\usepackage{enumitem}
\setlist{leftmargin=5.5mm}
\usepackage{makecell}
\usepackage{multirow}
\usepackage{lscape}
\usepackage{caption,subcaption}
\usepackage{svg}
\usepackage{hyperref}
\usepackage{xspace}
\usepackage{xcolor}
\usepackage{cellspace}
\usepackage[most]{tcolorbox}
\usepackage{booktabs,chemformula}
\usepackage{fixfoot}
\usepackage{balance}
\usepackage{tabularx}
\usepackage{float}
\usepackage{lscape}
\newcommand*{\ING}{ING\xspace}
\definecolor{darkblue}{RGB}{25, 43, 132}


\setcopyright{rightsretained}
\acmPrice{}
\acmDOI{10.1145/3611643.3616328}
\acmYear{2023}
\copyrightyear{2023}
\acmSubmissionID{fse23main-p838-p}
\acmISBN{979-8-4007-0327-0/23/12}
\acmConference[ESEC/FSE '23]{Proceedings of the 31st ACM Joint European Software Engineering Conference and Symposium on the Foundations of Software Engineering}{December 3--9, 2023}{San Francisco, CA, USA}
\acmBooktitle{Proceedings of the 31st ACM Joint European Software Engineering Conference and Symposium on the Foundations of Software Engineering (ESEC/FSE '23), December 3--9, 2023, San Francisco, CA, USA}
\received{2023-02-02}
\received[accepted]{2023-07-27}



\begin{document}

\title[Dynamic Prediction of Delays in Software Projects using Delay Patterns and Bayesian Modeling]{Dynamic Prediction of Delays in Software Projects using \\Delay Patterns and Bayesian Modeling}

\author{Elvan Kula}
\affiliation{%
  \institution{Delft University of Technology}
  \city{Delft}
  \country{The Netherlands}}
\email{e.kula@tudelft.nl}

\author{Eric Greuter}
\affiliation{%
  \institution{ING}
  \city{Amsterdam}
  \country{The Netherlands}}
\email{eric.greuter@ing.com}

\author{Arie van Deursen}
\affiliation{%
  \institution{Delft University of Technology}
  \city{Delft}
  \country{The Netherlands}}
\email{arie.vandeursen@tudelft.nl}

\author{Georgios Gousios}
\affiliation{%
  \institution{Delft University of Technology}
  \city{Delft}
  \country{The Netherlands}}
\email{g.gousios@tudelft.nl}

\renewcommand{\shortauthors}{Elvan Kula, Eric Greuter, Arie van Deursen, Georgios Gousios}

\begin{abstract} 
Modern agile software projects are subject to constant change, making it essential to re-asses overall delay risk throughout the project life cycle. Existing effort estimation models are static and not able to incorporate changes occurring during project execution. In this paper, we propose a dynamic model for continuously predicting overall delay using delay patterns and Bayesian modeling. The model incorporates the context of the project phase and learns from changes in team performance over time. We apply the approach to real-world data from 4,040 epics and 270 teams at \ING. An empirical evaluation of our approach and comparison to the state-of-the-art demonstrate significant improvements in predictive accuracy. The dynamic model consistently outperforms static approaches and the state-of-the-art, even during early project phases.
\end{abstract}

\maketitle

\input{sections/introduction}
\input{sections/RW}

\input{sections/context}

\input{sections/study-design}
\input{sections/patterns}
\input{sections/evaluation}
\input{sections/discussion}

\input{sections/threats-to-validity}

\input{sections/conclusion}
\input{sections/data-availability}

\bibliographystyle{acm}
\balance
\bibliography{mybib}

\end{document}

%% file: sections/introduction.tex
\section{Introduction}

\begin{figure}
\includegraphics[width=0.98\linewidth]{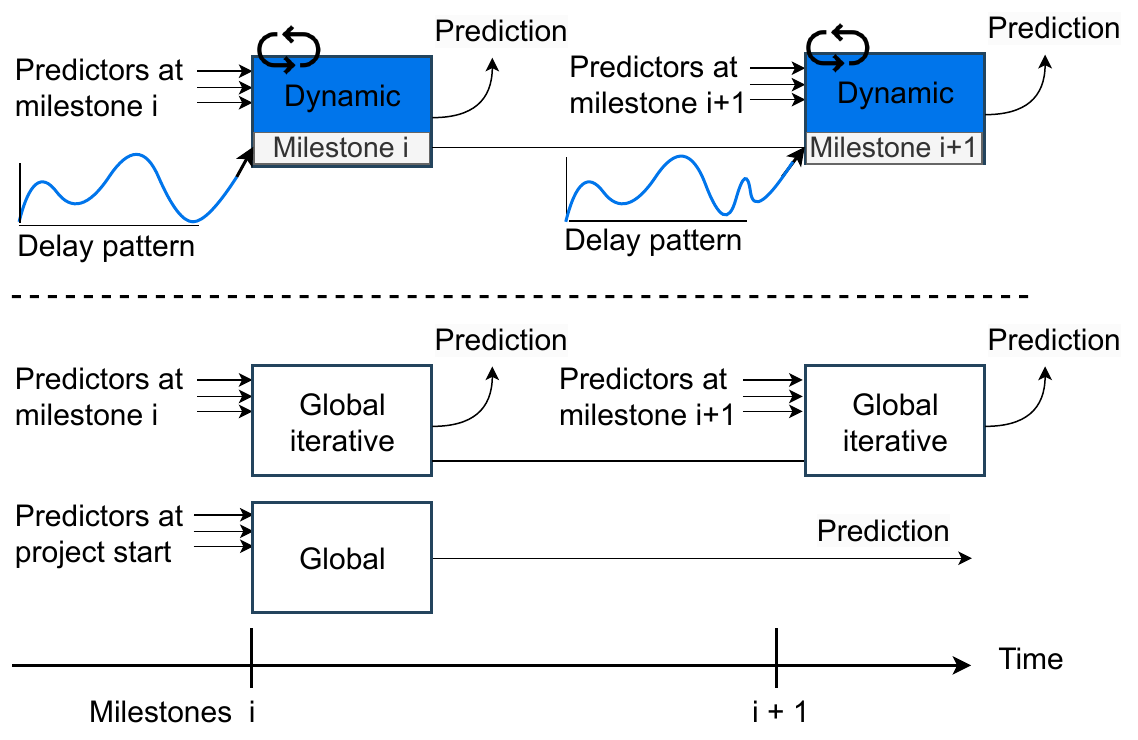}
\caption{Global, global iterative and dynamic approaches to delay prediction over time}
\label{model-diffs}
\end{figure}

Schedule delays constitute a major problem in the software industry. Software projects run, on average, around 30-40\% overtime~\cite{halkjelsvik2012origami, molokken2003review}. Ineffective risk management is one of the main reasons for delays in software projects~\cite{de2010does, han2007empirical}. An important activity involved in risk management is delay prediction. Foreseeing delay risks enables project managers to take measures to assess and manage risks, make timely adjustments to the planning and reduce delay propagation. \emph{Global} effort estimation models are the state-of-the-art in predicting overall delay for software projects~\cite{abrahamsson2007effort}. Global models are trained upfront and estimate the entire project using predictor variables collected at the beginning of the project. These models have a \emph{static character}: they capture the overall contribution of predictor variables to the total development effort and are unaware of changes occurring during project execution.

Global models are reasonable for traditional, waterfall-like settings where common predictors are known at the beginning of the project and do not change much throughout the project. However, this is not the case for modern, agile projects. In agile settings, projects (referred to as ``epics") are incrementally developed through short iterations to respond fast to changing markets and customer demands~\cite{cohn2005agile}. Predictors proposed in previous work~\cite{trendowicz2009factors, kula2021factors}, such as user requirements and task dependencies, can vary in value and relative impact during the execution of agile projects. Global models are not able to incorporate these changes due to their static character. An existing alternative is to use global models in an iterative manner (so-called \emph{global iterative})~\cite{abrahamsson2007effort}. That is, applying the global model at different prediction times throughout a project using updated predictor values. This may lead to an improvement in predictive accuracy. However, the global iterative model is still not able to adapt to changes occurring during project execution. Agile projects call for the need of models with a \emph{dynamic character}: models that are able to capture and adapt to changes in team performance and the impact of predictors during project execution.

In the field of transport, prediction of overall delay is an important requirement for proactive control of traffic and the feasibility of timetable realisation~\cite{corman2018stochastic}. Previous research in railway traffic~(e.g.,~\cite{artan2021exploring, huang2022enhancing}) and air transport~(e.g., \cite{oreschko2012turnaround, jiang2020applying}) has found that delays develop or propagate following certain patterns over time. A similar pattern in historic data can provide an estimation for the future development of delays. These studies detect patterns on the fly and use them for improving predictions of overall delay. It is not yet known whether this concept of delay patterns is applicable in the context of software development.

In this paper, we propose a \emph{dynamic} effort estimation model for continuously predicting overall delay in agile projects. As visualized in Figure~\ref{model-diffs}, the dynamic model extends global approaches by incorporating the context of the project phase (referred to as ``project milestone") and modeling delay patterns when making predictions. The dynamic model is updated after each milestone using the predictor values collected for that milestone and the development of delay up until that milestone. The model captures the milestone-specific contributions of predictors to the total development effort and follows changes in team performance over time.




To develop our dynamic model, we use a Bayesian modeling approach. Bayesian models are able to learn from changes in the relative impact of predictors by updating their beliefs. We train the Bayesian model on time series of predictors and intermediate delays recorded across the milestones of a project's timeline. Similar to prior work in transport, we apply time series clustering to identify recurrent delay patterns. We apply our dynamic approach to real-world data from 4,040 epics and 270 teams at \ING, a large Dutch internationally operating bank with more than 15,000 developers. We compare the performance of the dynamic Bayesian model with global approaches and the state-of-the-art baselines in software effort estimation.

An empirical evaluation of our approach demonstrates significant improvements in predictive performance, achieving on average 66--92\% Standardized Accuracy and 0.19--0.04 Mean Absolute Error over time. The dynamic model consistently outperforms global approaches and the state-of-the-art, even during early milestones (i.e., 10--30\% of project duration). The predictions of the dynamic model become substantially more certain and accurate over time.

The main contributions of this paper are:

\begin{itemize}[leftmargin=*,noitemsep,topsep=2pt]
    \item A new approach to predict delay using delay patterns and Bayesian modeling (Section 4)
    \item An application of the approach at \ING identifying four recurrent delay patterns (Section 5)
    \item An empirical evaluation of the approach and comparison to the state-of-the-art, clearly demonstrating a significant improvement in predictive accuracy (Section 6)
\end{itemize}


%% file: sections/RW.tex
\section{Related Work}

\textbf{Effort estimation models.} Prior work has been done in building models for estimating effort of the entire project (e.g., \cite{sarro2016multi, mair2000investigation, panda2015empirical}), a single iteration (e.g., \cite{choetkiertikul2017capability, abrahamsson2007effort, hearty2008predicting}) and a single software task~(e.g., a user story \cite{choetkiertikul2018deep, kula2021modeling} or issue report \cite{choetkiertikul2015threshold, choetkiertikul2017predicting, choetkiertikul2015predicting}). Existing models that estimate the total development effort are called \emph{global}~\cite{abrahamsson2007effort} and have a static character. They make a single prediction using predictors collected at the start of the development phase. Global models can be applied in an iterative manner to obtain estimates at different prediction times throughout development. Choetkiertikul et al.~\cite{choetkiertikul2017predicting} demonstrated this by applying their model for predicting delay risk at three different prediction times. They showed that the predictions become more accurate at later times since more information becomes available. Another study~\cite{choetkiertikul2015threshold} identified patterns of abnormal behaviors causing project delays and used these patterns to predict the delay risk of issues. The patterns are derived as combinations of threshold-exceeding risk factors that can lead to schedule overruns. 

\textbf{Effort drivers.} Previous research~\cite{trendowicz2014software} divided factors affecting the software development effort into four categories: personnel (i.e., team skills and experience \cite{mahnivc2012using, usman2014effort, torrecilla2015estimating, jorgensen2004review}), process (i.e., tools and methods used~\cite{abrahamsson2011predicting, nguyen2013review}), project (i.e., project management~\cite{kang2010model, inayat2015systematic}) and product characteristics (i.e., design and implementation~\cite{agrawal2007software}). Agile teams rely on documentation~\cite{pasuksmit2021towards} and expert judgement of team- and project-related factors~\cite{usman2015effort, dantas2018effort} for estimation of software tasks and iterations. Kula et al.~\cite{kula2021factors} identified the most relevant factors and their interactions affecting the effort of epics.

\textbf{Delay patterns in transport.} Previous research in railway traffic~(e.g., \cite{artan2021exploring, huang2022enhancing}) and air transport~(e.g., \cite{oreschko2012turnaround, jiang2020applying}) has shown that delays develop or propagate following recurrent patterns over time. These patterns can provide information on the future development of delays. Artan and Sahin~\cite{artan2021exploring} used Markov chains to model patterns of delay deterioration, recovery and state keeping in train running times. Huang et al.~\cite{huang2022enhancing} used a clustering technique to identify four types of delay patterns in train operations: decreasing delays, unchanged delays, small increasing delays and large increasing delays. They built a Bayesian Network model that uses the patterns in previous train stations to predict delay for upcoming stations. Oreschko et al.~\cite{oreschko2012turnaround} detected specific delay patterns in flight arrival times with respect to the time of day and airport category. Jiang et al.~\cite{jiang2020applying} uses patterns of flight delay as input for a machine learning-based approach to delay prediction.

While delay patterns have been proven useful for delay prediction in transport, they remain unexplored in the context of software development. It is unclear whether and how delay patterns can be employed in software projects. Our study complements prior work by modeling delay patterns and using them as input for a dynamic approach to predict overall delay in software projects.

%% file: sections/context.tex
\section{Background}\label{background}


\textbf{Agile software development.}
In agile settings, user requirements are expressed based on Leffingwell's five-level hierarchy~\cite{leffingwell2007scaling}. The \emph{strategic themes} of a company are divided into \emph{epics} that represent high-level product requirements~\cite{cohn2005agile}. Epics are large pieces of functionality that can be split into \emph{features}, which in turn can be split into \emph{user stories}. Stories are short requirements written from the perspective of an end user~\cite{cohn2004user}. Agile teams use a product backlog to keep track of the status and priority of these work items~\cite{schwaber2002agile}.

Agile teams start with a high-level release plan (typically 2-6 months) that centers on epics~\cite{cohn2004user}, which encompass multiple teams across multiple iterations. An iteration is a fixed, short period of time (typically 1-4 weeks) in which a single development team delivers a set of user stories. Iteration planning focuses on selecting and estimating the user stories that need to be delivered in the next iteration. Agile teams rely on expert judgement to estimate the effort of a user story in story points~\cite{usman2015effort, cohn2005agile}. They usually do this in structured group meetings (e.g., using Planning Poker~\cite{grenning2002planning}). At the end of every iteration, teams review which user stories are completed and which ones need to be delayed to the next iteration. These progress updates can be leveraged to refine release/epic plans during execution. 


\textbf{Epics at \ING.}
We performed an evaluation of our dynamic prediction approach at \ING TECH, the IT department that is responsible for the main banking applications used by millions of customers. The department has significant variety in terms of products, size and application domain. 
Teams at \ING follow Scrum~\cite{cervone2011understanding} as agile methodology and work in iterations of 1 to 4 weeks. They usually deliver epics in a time span of three to 12 months. Developers use Planning Poker~\cite{grenning2002planning} and a fixed Fibonacci sequence of values for estimating story points. The teams estimate user stories in working hours and then convert them into story points. The rule of thumb at \ING TECH is that a one-point story should take about half a day of work (4 hours). A two-point story should be twice as much effort, that is, one day of work (8 hours). This makes the story points additive and comparable between teams.



\textbf{Bayesian data analysis.} Recent works~\cite{furia2019bayesian, furia2022applying, torkar2021method} identified the potential of Bayesian statistical techniques in software engineering research. Bayesian models are flexible, easy to interpret and provide a detailed probability distribution~\cite{furia2019bayesian}. They are based on a uniform framework that applies Bayes' theorem to update prior beliefs about model parameters based on observed data. Bayesian models consist of three components~\cite{mcelreath2020statistical}:

\begin{itemize}[leftmargin=*,noitemsep,topsep=3pt]
    \item \emph{Likelihood}: A function that represents the probability of observing the data given a set of model parameters. It reflects the underlying data generation process. In the context of delay prediction, the likelihood captures the probability of observing a specific delay value or a set of delay values.
    \item \emph{Priors}: Probability distributions that represent the initial beliefs or assumptions about the model parameters before observing the data. Priors allow incorporating existing knowledge about the effects of predictors.
    \item \emph{Posterior}: The updated probability distribution that incorporates both the prior information and the likelihood of the observed data. It is obtained by repeatedly sampling values from the priors and applying Bayes' theorem using the likelihood. The posterior is used to make predictions about future observations.
\end{itemize}

%% file: sections/study-design.tex
\section{Approach}
Our overall research goal is to extract delay patterns and build a dynamic model that incorporates the patterns and the context of the project phase for continuously predicting the overall delay of an epic. This requires dividing an epic's timeline into designated milestones (Section~\ref{sec:milestones}) and tracking of intermediate delay and predictors across these milestones. The milestones should match the work pace of the organization and can be set accordingly at fixed time intervals or fractions of the planned project duration. It is a very common practice of agile teams to record the delivery status of their work items in a backlog management tool. Backlog data can be used to extract intermediate delay and predictors over milestones in the form of time series (Section~\ref{sec:data-collection}). 

To identify delay patterns, the time series of delay values over milestones need to be partitioned into groups of similar elements using clustering (Section~\ref{clustering-model}). Hierarchical clustering or K-means can be used to identify and discriminate different recurrent patterns. The dynamic prediction model is learned using the time series data of the clustering output and predictor values (Section~\ref{sec:diff-modes}). For the Bayesian model, it is important to select the likelihood and tune the priors based on the dataset being used (Section~\ref{sec:bayesian-dev}). At each milestone, the updated variables are fed into the model to obtain a new, refined estimate and update the model's beliefs. This way the model learns and evolves with the epic over time.

\subsection{A Unified Timeline of Project Milestones}\label{sec:milestones} 
To incorporate the context of the project phase, we present the timeline of an epic delivery as a sequence of regularly-spaced milestones. It is important to use a unified timeline so that delay values measured at the milestones can be aggregated across epics for pattern identification. Since teams working on an epic can follow different iteration lengths, we cannot use iterations or fixed time intervals. Instead, we define the milestones based on completion rate to evenly space them out along deliveries. The completion rate is based on the number of iterations completed compared to the total number of iterations planned. For example, an epic that consists of 20 iterations will achieve its 10\% milestone after completing the initial two iterations. The total number of milestones used will determine the granularity of the collected time series and, therefore, the identified patterns. As progress updates are given at the end of every iteration, target milestones that cover the iteration length used (usually 2-4 weeks~\cite{cervone2011understanding}). 

\textbf{Milestones at \ING.} The average iteration length in our dataset at \ING is 16 days. We performed our analysis with 10 milestones, which breaks most epics at \ING down into intervals of two to three weeks with an average duration of 17 days. In total, 17\% of the epics at \ING consist of less than 10 iterations; we excluded those from our dataset to keep only the epics that have at least one iteration update available at every milestone (see Section~\ref{sec:pre-process}). Each epic is divided into 10 milestones: every milestone is scored as 10\% of the planned duration, so when a team reaches the third milestone of a task, their completion rate is equal to 30\% and so on. When an epic's total number of iterations is not divisible by 10, we round the milestones off to the last completed iteration of their time frame. For example, when an epic consists of 18 iterations, its sixth milestone ($\frac{6}{10} \times 18 = 10.8$) will be measured at the end of the $10^{th}$ iteration. The milestones are connected by the corresponding iterations and occur in a fixed sequence $j \rightarrow k$, where $k = j + 1$, $j = 1, 2, ..., 10$.

\subsection{Data Collection}\label{sec:data-collection}
\subsubsection{Backlog data}
To track changes in the intermediate delay and predictors over time, we need a backlog dataset that contains the history of epics. For each epic, this dataset has to include the \emph{identification number}, \emph{creation date}, \emph{planned start date}, \emph{actual start date}, \emph{planned delivery date} and \emph{actual delivery date}. At \ING, we extracted log data from the backlog management tool \emph{ServiceNow}. This dataset contained 7,463 epics delivered by 418 teams between January 01, 2017 and January 01, 2022. 

\subsubsection{Data cleaning}\label{sec:pre-process} 
To eliminate noise and missing values, epics with a status other than `Completed' need to be removed. Epics that are not assigned to any team or have empty \emph{Planned Delivery Date} and \emph{Actual Delivery Date} fields also need to be filtered out. At \ING, we chose to exclude the epics that consist of less than 10 sprints to keep only the epics that have at least one sprint update available at every milestone. In addition, we removed epics that exceed two standard deviations from the mean overall schedule delay of all epics. After linking and cleaning the data, the final \ING dataset was reduced to 4,040 epics from 270 teams.

\subsubsection{Delay factors} The predictor variables can be obtained over milestones by extracting their values at the end of the last iteration that corresponds to a milestone. We extracted 13 predictor variables that represent factors affecting delays in epic deliveries. We identified these factors in previous work~\cite{kula2021factors}. We used the same procedure to extract the predictor variables. Table~\ref{tab:proxies} provides an overview of the predictors we collected and the influential factors they correspond to. For example, we model the delay factor \emph{team familiarity} using the predictor variable \emph{team-existence} that measures the amount of time team members have worked together in their current composition.



\subsubsection{Measuring schedule deviation}\label{sec:bre} 
We measure the overall schedule delay at the end of an epic using \emph{Balanced Relative Error}~(BRE)~\cite{miyazaki1994robust} as error measure. BRE has been recommended as an unbiased alternative to the commonly used Mean of Magnitude of Relative Error and Prediction at level l \cite{foss2003simulation, kitchenham2001accuracy, port2008comparative}. BRE is defined as:

\begin{equation*}
\textrm{If Act - Est} \geq 0, \textrm{then BRE} = \frac{\textrm{Act - Est}}{\textrm{Planned duration}} 
\end{equation*}

\begin{equation*}
\textrm{If Act - Est} < 0, \textrm{then BRE} = \frac{\textrm{Act - Est}}{\textrm{Actual duration}} 
\end{equation*}
\vspace{0.8em}

where $Act$ is the actual delivery date and $Est$ is the planned delivery date of an epic. $Act - Est$ calculates the schedule deviation in days: a positive difference corresponds to a delay ($Act$ is later than $Est$) and a negative value corresponds to on-time delivery ($Act$ is before $Est$). $\textit{Actual duration}$ is the difference (in days) between the actual delivery date and start date of an epic. $\textit{Planned duration}$ is the difference (in days) between the planned delivery date and start date of an epic. 

To measure the intermediate delay of an epic at a given milestone, we select the last iteration corresponding to that milestone and calculate the total number of story points that are delayed to the next iteration/milestone. The total number of delayed story points represents the workload of user stories a team was unable to complete or resolve.

\begin{table*}[!tbp]
\footnotesize
\centering
\caption{The 13 extracted predictor variables representing factors from \cite{kula2021factors} that affect delays in epic deliveries. The \emph{Description} column provides a description of each variable.}
\label{tab:proxies}
\begin{tabular}{p{2.7cm} | p{2.7cm} p{8cm}}
\toprule
\textbf{Risk factor} & \textbf{Predictor variable} & \textbf{Description}\\
\midrule
Task dependencies & 1. out-degree & Number of outgoing dependencies of an epic on other epics \\ 

Organizational stability & 2. changed-leads & Number of changed tribe leads during the current and previous epic \\ 

Team stability & 3. stability-ratio & Median of the ratio of team members that did not change during the current and previous epic\\[1pt] 

Skills and knowledge & 4. dev-age-ing & Median of the number of years the developers working on the epic have been working at \ING \\[1pt]

Team familiarity & 5. team-existence & Median of the number of years teams have existed in their current composition of team members\\ 

Team commitment & 6. hist-performance & Median of the ratio of on-time delivered epics over all teams working on the epic\\ 

Work in progress & 7. dev-workload & Median of the number of story points assigned to a developer per sprint \\ 

Bugs or incidents & 8. nr-incidents & Number of incidents that occurred during the development phase of the epic \\[3pt]

 & 9. unplanned-stories & Number of unplanned stories (related to bug fixes or incidents) that have been added to the epic \\[1pt] 
                 
Project size & 10. nr-stories & Number of planned stories assigned to the epic \\[1pt] 

& 11. nr-sprints & Number of sprints assigned to the epic \\[1pt] 

& 12. team-size & Median team size in the epic\\[1pt]
Project security & 13. security-level & The ratio of user stories in the epic that need to pass a security testing process \\ 
    \bottomrule
    \end{tabular}
\end{table*}

\subsection{Clustering for Delay Pattern Discovery}\label{clustering-model}
To identify delay patterns, the time series of intermediate delay values recorded across milestones need to be clustered into groups of similar elements. In agile settings, intermediate delay can be measured based on the number of delayed user stories or story points. We measure the number of \emph{Delayed Story Points} (DSP) as it is a more specific measure of the delayed workload. We calculate the delayed story points at a given milestone~$i$ as the number of story points that are delayed to the next milestone~$i + 1$. DSP represents the delayed workload at a particular milestone and is thus not cumulative. We normalize the DSP values per epic to make sure that the range of story point values cannot influence the clustering results. We divide the DSP values by the maximum number of story points that are delayed along the timeline of an epic. 

\begin{figure}[t!]
    \centering 
  \includegraphics[width=0.90\linewidth]{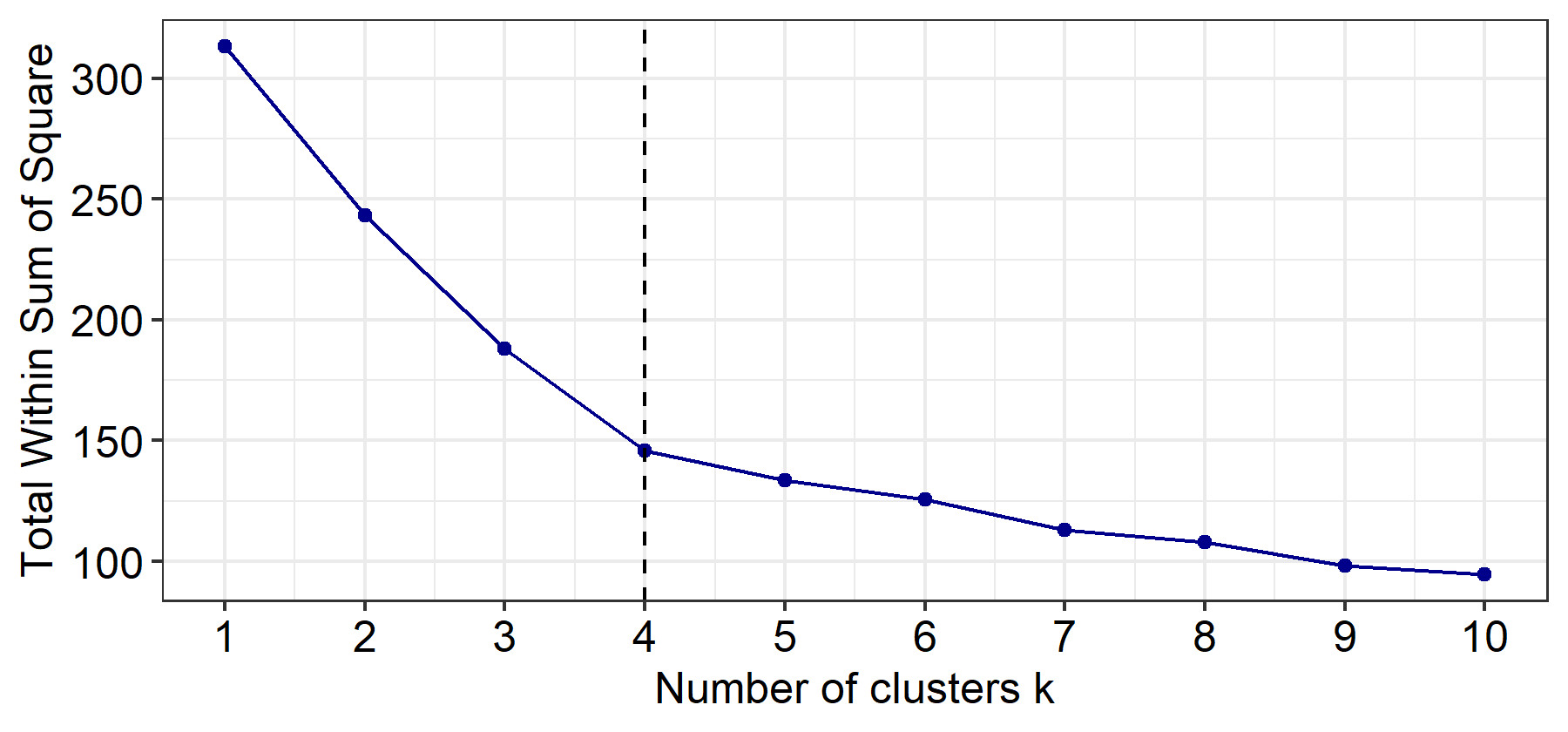}
  \caption{Elbow method and WSS curve for selecting the optimal number of clusters}
  \label{fig:elbow}
\end{figure}

To partition the time series data, we use hierarchical clustering with \emph{Dynamic Time Warping} (DTW) as distance measure~\cite{muller2007dynamic}. This approach has been shown to be appropriate for short time series~\cite{aghabozorgi2015time}. DTW is a shape-based distance measure that finds optimal alignment between two time series that do not necessarily match in time or length. This makes DTW particularly suitable for epics that can differ in duration and sprint length, which is the case at \ING. We use the Elbow method to select the optimal number of clusters \emph{k}. This method calculates the total Within Cluster Sum of Squares (WSS)~\cite{hartigan1979k} for each k. The point of inflection on the WSS curve indicates the optimal number of clusters. Figure~\ref{fig:elbow} presents the WSS curve for \ING data and shows that a $k$ value of $4$ is optimal. 


The application of our clustering approach to \ING data resulted in four patterns that are discussed in Section~\ref{sec:patterns}. To characterize the clusters in terms of risk factors, we applied the Wilcoxon Signed Rank Test~\cite{arcuri2014hitchhiker} for pairwise comparisons. This is a non-parametric test that makes no assumptions about underlying data distributions.

\subsection{Bayesian Model Development}\label{model-dev}


The main goal of our prediction model is to infer a probability distribution of BRE values across milestones. We use Bayesian statistical analysis to infer the probabilities and build the model in global and dynamic modes for comparison.

\subsubsection{Different modes of model development}\label{sec:diff-modes}
We build and compare the Bayesian model using global, global iterative and dynamic modes of development. The differences between the models are visualized in Figure~\ref{model-diffs} and can be explained as follows:

\begin{itemize}[leftmargin=*,noitemsep,topsep=2pt]
\item The \textbf{global} model solely uses the predictor variables as input and does not have a sense of time. It makes a single prediction of the overall delay based on predictors collected at the start of the project and does not update its BRE estimate throughout the project.

\item The \textbf{global iterative} model is the global model used in an iterative manner (i.e., over milestones). We apply the global model at each milestone to obtain a new estimate of the overall delay based on the predictor values of that milestone. The model itself is not updated.

\item The \textbf{dynamic} model is learned using the time series data of the clustering output and predictor values collected over milestones. This model incorporates the context of the milestone and thus has a sense of time. At each milestone~$i$, the clustering model classifies the set of delay values across previous milestones $1$ to $i-1$ into one of the four identified groups of patterns (producing a pattern label). To mimic a real prediction scenario, we set the values for future milestones $i+1$ to $n$ to zero (unknown) in the input data for the clustering model. At each milestone, the pattern label and updated predictor variables are fed into the dynamic model to obtain a new estimate of the overall delay and update the model's posterior distribution. 
\end{itemize}

\subsubsection{Bayesian modeling}\label{sec:bayesian-dev}
We use Bayesian regression analysis to infer the probabilities that quantify delay risk and propagate uncertainty over time. We implemented our models in the statistical framework \texttt{Stan}\footnote{https://mc-stan.org/}. We designed the models following the steps and guidelines for Bayesian data analysis in software engineering research~\cite{furia2019bayesian, furia2022applying, torkar2021method}:

\emph{Step 1. Selecting a likelihood.} The choice of a likelihood function depends on the type of data. The BRE values are proportional numbers between 0 and 1. In total, 42\% of the BRE values in the \ING dataset are zero (corresponding to on-time delivered epics). The data does not contain BRE values of one; the maximum BRE in our dataset is 0.83. A common choice for modeling proportional data is the Beta distribution likelihood~\cite{ferrari2004beta}. Beta models are highly flexible and can take on all sorts of different shapes. To account for the zero values in the \ING dataset, we selected the Zero-Inflated Beta distribution~\cite{ospina2012general}, relating predictors to outcome, as shown in Eq.~\ref{eq:1}. The Zero-Inflated Beta distribution depends on a mean $\mu$ and precision $\phi$, like in a regular Beta, but it may produce a BRE of zero with probability $\alpha$ in each draw from the distribution. We used a logit function for $\mu$ and $\alpha$ to translate them back to the log-odds scale of the (0,1) scale. We assume that all predictor variables may affect the parameters of the model (Eq.~2--4). \\ 

\begin{figure}[t!]
\begin{subfigure}[h]{0.48\linewidth}
\includegraphics[width=\linewidth]{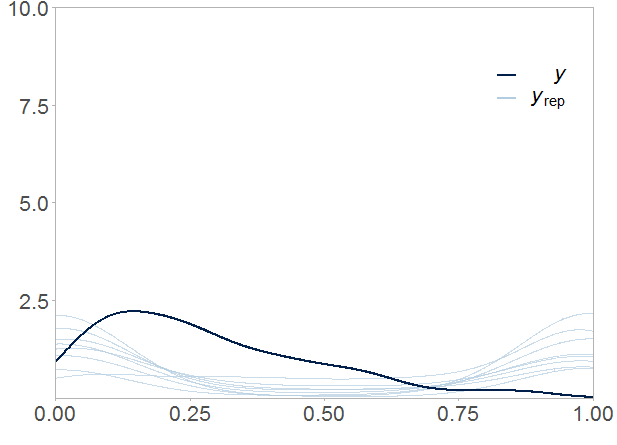}
\caption{Prior predictive check}
\label{fig:prior-check}
\end{subfigure}
\hfill
\begin{subfigure}[h]{0.48\linewidth}
\includegraphics[width=\linewidth]{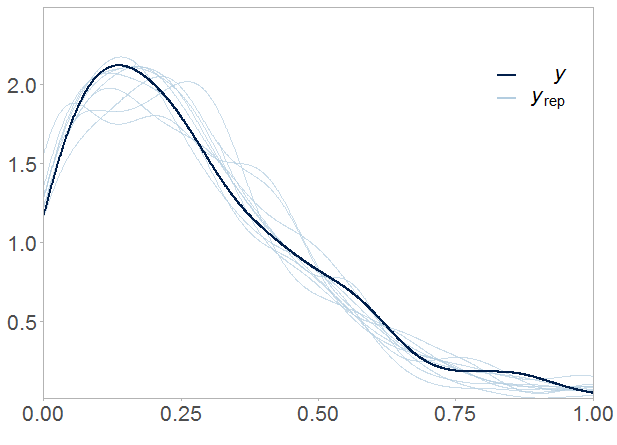}
\caption{Posterior predictive check}
\label{fig:posterior-check}
\end{subfigure}%
\caption{Density overlays of predictive prior and posterior draws (visualized as light blue lines) versus the real data (shown as the dark blue line). The combination of our priors (left plot) shows that we assign more probability mass to low and high BRE values. After making use of the data (right plot) we see a good model fit: the light blue lines are covering the dark blue line.} 
\label{model_checks}
\end{figure}

\emph{Step 2. Setting priors.} To apply Bayes' theorem, we need to define priors for the model's parameters. A common approach, which works well in most cases, is a weakly informative prior~\cite{lemoine2019moving}, such as a normal distribution with zero mean and moderate standard deviation, as shown in Eq. 5 and 7. Such a prior does not bias the effect that the predictors may have towards positive or negative values, and it still allows for a wide range of parameter values. We set a Cauchy distribution (Eq. 6) for the $\beta_{\phi}$ parameters, which is a common choice for dispersion parameters~\cite{gelman2006prior}. To check what the combination of priors implies on our outcome, we sample from the priors only. This is called \emph{prior predictive checks} (see Figure~\ref{fig:prior-check}).\\

The overall definition of the dynamic model is given in Eq. 1--7.  
\setlength{\abovedisplayskip}{4pt}
\setlength{\belowdisplayskip}{4pt}
\begin{align}
\begin{split}\label{eq:1}
    \text{BRE}_{i} \sim{}& \text{Zero-Inflated Beta}(\mu_i, \phi_i, \alpha_{i})
\end{split}\\
\begin{split}\label{eq:2}
\text{logit}(\mu_i) ={}& \beta_{\mu_1} \cdot \text{out-degree} + \ldots + \beta_{\mu_{13}} \cdot \text{security-level} \\
+{}& \beta_{\mu_{14}} \cdot \text{milestone} + \beta_{\mu_{15}} \cdot \text{DSP}  + \beta_{\mu_{16}} \cdot \text{pattern}
\end{split}\\
\begin{split}\label{eq:3}
\text{log}(\phi_{i}) ={}& \beta_{\phi_1} \cdot \text{out-degree} + \ldots + \beta_{\phi_{13}} \cdot \text{security-level} \\
+{}& \beta_{\phi_{14}} \cdot \text{milestone} + \beta_{\phi_{15}} \cdot \text{DSP}  + \beta_{\phi_{16}} \cdot \text{pattern}
\end{split}\\
\begin{split}\label{eq:4}
\text{logit}(\alpha_i) ={}& \beta_{\alpha_1} \cdot \text{outdegree} + \ldots + \beta_{\alpha_{13}} \cdot \text{security-level}\\
+{}& \beta_{\alpha_{14}} \cdot \text{milestone} + \beta_{\alpha_{15}} \cdot \text{DSP}  + \beta_{\alpha_{16}} \cdot \text{pattern}
\end{split}\\
\begin{split}\label{eq:5}
\beta_{\mu_{1}}, \ldots, \beta_{\mu_{16}} \sim{}& \text{Normal}(0,1)
\end{split}\\
\begin{split}\label{eq:6}
\beta_{\phi_{1}}, \ldots, \beta_{\phi_{16}} \sim{}& \text{Cauchy}(0,1)
\end{split}\\
\begin{split}\label{eq:7}
\beta_{\alpha_{1}}, \ldots, \beta_{\alpha_{16}} \sim{}& \text{Normal}(0,1)
\end{split}
\end{align}

The `pattern' predictor in Eq. 2--4 stands for the delay pattern label as classified by the clustering model. The global and global iterative models follow the same design, except that they do not include the milestone and pattern label as predictors. \\

\emph{Step 3. Sampling.} For sampling, we used the Hamiltonian Monte Carlo implementation that Stan provides. To improve the efficiency of sampling, we centered and scaled all predictor variables. Once the model has been sampled, we check diagnostics to ensure that we have reached a stationary posterior distribution. No warnings regarding divergent transitions and low E-BFMI values were reported~\cite{betancourt2017conceptual}. Moreover, the $\hat{R}$ diagnostic was consistently less than 1.01 and the ESS value was higher than 0.2. This indicates that the Markov chains mixed well~\cite{vehtari2021rank}. To check if the model fits the data, we sample from the priors with data. This is called \emph{posterior predictive checks} (see Figure~\ref{fig:posterior-check}). A summary of the model can be found in the supplemental material~\cite{supplemental}. On the 95\% level, all predictors have a significant effect. \\

\emph{Step 4. Model checking.} To check for overfitting, we test whether any model making simpler assumptions about the data performs comparably or better than our model with the Zero-Inflated Beta distribution ($M_{ZIB}$). We compare $M_{ZIB}$ with simpler models in terms of expected log predictive
density (ELPD) using leave-future-out cross-validation~\cite{burkner2020approximate, vehtari2017practical}. The models are conditioned on two years of historical data (covering the epics from 2017 to 2019) using the recommended threshold of 0.7 for the Pareto $k$ estimates~\cite{vehtari2017practical}. The results of our analysis can be found in the supplemental material~\cite{supplemental}. The results show that $M_{ZIB}$ performs significantly better than other, simpler models and thus fits the data better while avoiding overfitting.

%% file: sections/patterns.tex
\section{Delay Patterns at \ING}\label{sec:patterns}

\begin{figure}[t!]
    \centering 
  \includegraphics[width=\linewidth]{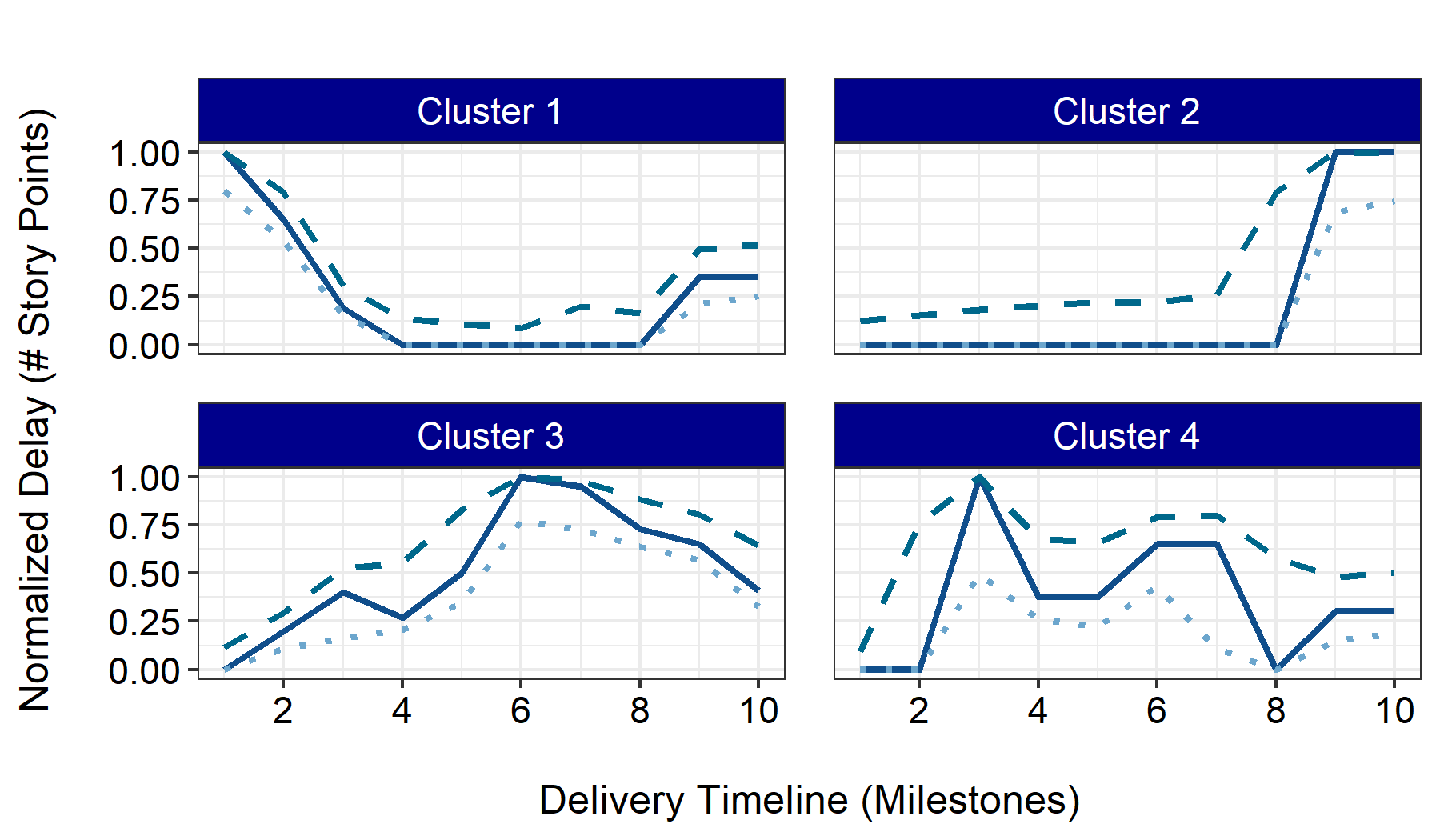}
  \caption{Four clusters of delay profiles representing recurrent delay patterns across milestones in epic deliveries at \ING: 25th percentile: dotted; centroid: solid; and 75th percentile: dashed.}
  \label{fig:patterns}
\end{figure}

\begin{table}\centering
\small
\noindent
\begin{tabular}{lllllcccc}
\toprule
\multirow{2}[3]{*}{\textbf{Predictor}} & \multicolumn{4}{c}{\textbf{Median}} & \multicolumn{4}{c}{\textbf{Significance}} \\
\cmidrule(lr){2-5} \cmidrule(lr){6-9}
 &  C1 & C2 & C3 & C4 & C1 & C2 & C3 & C4  \\
\cmidrule(lr){2-5} \cmidrule(lr){6-9}
nr-sprints  & 13 & 15 & 14 & 11 & & &  & $\ast$ \\
out-degree  & 7 & 3 & 4 & 4 & $\ast$ & & &  \\
hist-performance & 0.69 & 0.67 & 0.74 & 0.61 & & & & \\
dev-age-ing  & 2.49 & 2.61 & 2.92 & 2.84  & &  &  &  \\
team-existence  & 1.30 & 1.53 & 1.29 & 1.42 & &  & &   \\
team-size & 8 & 7 & 6 & 7 & & & $\ast$ & \\
security-level & 0.56 & 0.77 & 0.53 & 0.36 & & $\ast$ & & $\ast$  \\
unplanned-stories & 0.11 & 0.16 & 0.10 & 0.08  & & $\ast$ & & \\
changed-leads  & 3 & 2 & 3 & 2  & & & & $\ast$  \\
stability-ratio & 0.73 & 0.81 & 0.64 & 0.72  & & $\ast$ & &  \\
nr-stories & 52 & 43 & 39 & 45 & $\ast$ & & & \\
nr-incidents & 8 & 12 & 8 & 6  & & $\ast$ & &  \\
dev-workload & 15 & 12 & 10 & 8 & $\ast$ & & & $\ast$  \\
\cmidrule(lr){1-9}
BRE & 0.23 & 0.17 & 0.11 & 0.09 & $\ast$ & $\ast$ & $\ast$ & $\ast$ \\
\bottomrule
\end{tabular}
  \caption{Characteristics of delay profile clusters: Cluster 1 (C1), Cluster 2 (C2), Cluster 3 (C3), Cluster 4 (C4). $\ast$ indicates that a cluster is significantly different from all other clusters for the corresponding predictor variable (pairwise Wilcoxon tests with Bonferroni correction).}
  \label{tab:rq1-2-characteristics}
  \vspace{-1.5em}
\end{table}

Using the Elbow method, we determined $k=4$ as the optimal number of clusters (see Fig.~\ref{fig:elbow}). Therefore, we obtained four clusters representing delay patterns in epics at \ING. Figure~\ref{fig:patterns} visualizes the centroids and the 25th and 75th percentiles of the cluster delay distributions. The epics are grouped together with low mean variance (Var) around the cluster centroids (Var $C1 = 0.07$, Var $C2 = 0.11$, Var $C3 = 0.08$, Var $C4 = 0.12$), highlighting recurrent patterns. \newline

\textbf{Characteristics of clusters.} Table~\ref{tab:rq1-2-characteristics} summarizes the statistics of the predictor variables for each cluster's epics. The confidence intervals are included in the supplemental material~\cite{supplemental}. We used the Wilcoxon test for pairwise comparisons (Bonferroni corrected) to identify the factors for which clusters are significantly different from the other three clusters (highlighted with an $\ast$ in Table~\ref{tab:rq1-2-characteristics}). These factors characterize the epics exhibiting one of the four recurrent patterns. Even though we cannot reason about causal links between the factors and patterns, the results of the analysis enable us to form hypotheses on the causes of delays. Testing such hypotheses could lead to actionable insights and suggest delay mitigation measures. The clusters can be described as follows:

\begin{itemize}\setlength\itemsep{0.7em}
    \item \emph{Cluster 1} (C1) consists of 1388 (36\%) epics. These deliveries start out with a delay peak, followed by multi-phase recovery, and end with delay that continues beyond the planned delivery date. The epics of C1 have a significantly higher number of outgoing dependencies and developer workload, likely causing issues at the start of the delivery.
    \item \emph{Cluster 2} (C2) makes up the largest group, containing 1706 (44\%) epics that are punctual up until the last few milestones. The epics of C2 have a significantly higher security level and team stability, possibly explaining the consistent start. They also run into more incidents and unplanned work, likely causing the delay at the end of the delivery.
    \item \emph{Cluster 3} (C3) contains 540 (14\%) epics that exhibit an upward trend (i.e., delay increase) in the first section of the delivery followed by resilient recovery. The epics of C3 involve significantly smaller teams, suggesting that teams with fewer members may need some buildup time to respond to delay.
    \item \emph{Cluster 4} (C4) contains 232 (6\%) epics that exhibit a fluctuating pattern of delay increase and recovery over the course of the delivery. The epics of C4 have a significantly higher stability and lower security level, developer workload and number of sprints. These characteristics might possibly explain the consistent recovery of delay over time.
\end{itemize}
\vspace{0.5em}
\textbf{Patterns are indicative of overall delay.} The bottom row of Table~\ref{tab:rq1-2-characteristics} provides the descriptive statistics of the overall delay, measured in terms of BRE values, for each cluster. The epics assigned to Cluster 1 suffer the largest overall delay with a median BRE of 0.23. The epics in Cluster 2 are associated with the second largest overall delay (median BRE of 0.17). Clusters 3 and 4 consist of epics that end up with small overall delays with a median BRE of 0.11 and 0.09, respectively. Using the the Wilcoxon test for pairwise comparisons, we found that the differences in the BRE values of the clusters are statistically significant at a confidence level of 95\%. This means that the patterns are indicative of the overall epic delay.


%% file: sections/evaluation.tex
\section{Evaluation}

\subsection{Research Questions}

The empirical evaluation of the dynamic model aimed to answer the following research questions:

\begin{itemize}[label={--}]
\setlength\itemsep{0.6em}
\item \textbf{RQ1. Benefits of dynamic prediction:} \emph{Does the dynamic model provide more accurate estimates than its global and global iterative modes?} To study the benefits of the proposed dynamic model, we evaluate the performance of the Bayesian model in global, global iterative and dynamic settings.

\item \textbf{RQ2. Benefits of delay patterns:} \emph{Does the use of delay patterns have a positive impact on the predictive performance?} We compare the performance of the dynamic Bayesian model learned with and without the delay patterns.

\item \textbf{RQ3. Comparison with SoTA baselines:} \emph{How does our dynamic Bayesian model compare to the state-of-the-art baselines?} To determine whether our dynamic Bayesian model improves the state-of-the-art (SoTA) baselines in effort estimation, we compare it with the Decision Tree model of Choetkiertikul et al.~\cite{choetkiertikul2015threshold} and the Random Forests model of Choetkiertikul et al.~\cite{choetkiertikul2017predicting}. We perform the comparison with the models in their original, global mode using features from Choetkiertikul et al. and in dynamic mode using our set of features.

\item \textbf{RQ4. Impact of prediction time:} \emph{How does the moment of prediction affect the informativeness of the predictions of the dynamic model?} Previous work~\cite{jorgensen2021evaluation, jorgensen2019evaluating} has shown that statistical models should be evaluated in terms of both accuracy and informativeness (i.e., width of the prediction interval). We analyze how the informativeness of the predictions of the dynamic model evolves with the time of prediction (early versus late in the epic).
\end{itemize}

\subsection{SoTA Baselines}\label{baseline-models-dev}
We implemented two models representing the SoTA baselines in their original, global mode and dynamic mode for comparison with our dynamic Bayesian model. For comparison in global mode, we implemented the global Decision Tree model of Choetkiertikul et al.~\cite{choetkiertikul2015threshold} using the five issue-level features presented in the paper. We mapped the features to the epic-level and extracted them from \ING data. An overview of all variables and their mapping to the epic-level can be found in the supplemental material~\cite{supplemental}. We also implemented the global iterative Random Forests model of Choetkiertikul et al.~\cite{choetkiertikul2017predicting} using 16 out of 19 features from the paper. We were not able to extract the variables `number of fix versions', `changing of fix versions' and `number of affect versions' as they are specific to the context of issue reports. We converted the features to the epic-level, as described in the supplemental material. For comparison in dynamic mode, we implemented both models of Choetkiertikul et al. following the dynamic setup described in Section~\ref{sec:diff-modes}. The models were learned using our features from Table~\ref{tab:proxies} and the delay patterns.


\subsection{Experimental Setup}\label{model-evaluation}
We performed experiments on the 4,040 epics in the \ING dataset. To mimic a real prediction scenario, in which observed epics are used to inform predictions for future epics, we sorted the epics and their milestones based on their start date. For training and evaluation, we used time-based 10-fold cross-validation. The time-based variant of cross-validation ensures that in the k-th split, the epics in the first k folds (training set) are created before the epics in the (k+1)th fold (test set). The successive training sets are thus supersets of previous ones. This allows for the sequential updating of models based on past knowledge. 

The Bayesian model estimates a probability distribution of BRE values. For evaluation, we selected the median of the posterior distribution as the predicted BRE value. This is a common approach when the goal of the model is to minimize the absolute or relative estimation error~\cite{jorgensen2019evaluating}. For the SoTA baselines, we applied the Decision Tree and Random Forests regressors to obtain a BRE estimate. 


\begin{figure*}[!t]
        \centering
                \begin{subfigure}[b]{0.485\linewidth}  
        \centering 
        \includegraphics[width=\textwidth]{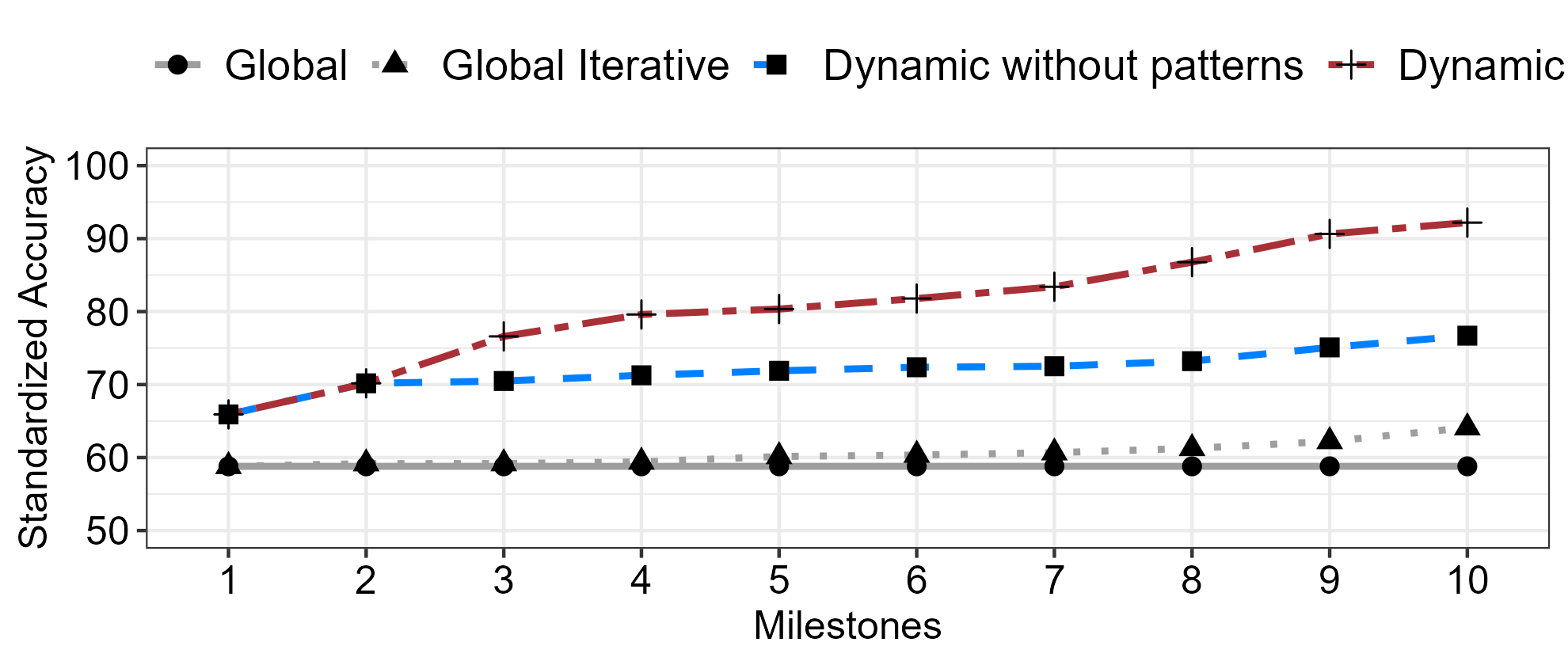}
            \caption{Standardized Accuracy over time}
            \label{fig:line-sa}
        \end{subfigure}
        \hfill
        \begin{subfigure}[b]{0.485\linewidth}
            \centering
            \includegraphics[width=\textwidth]{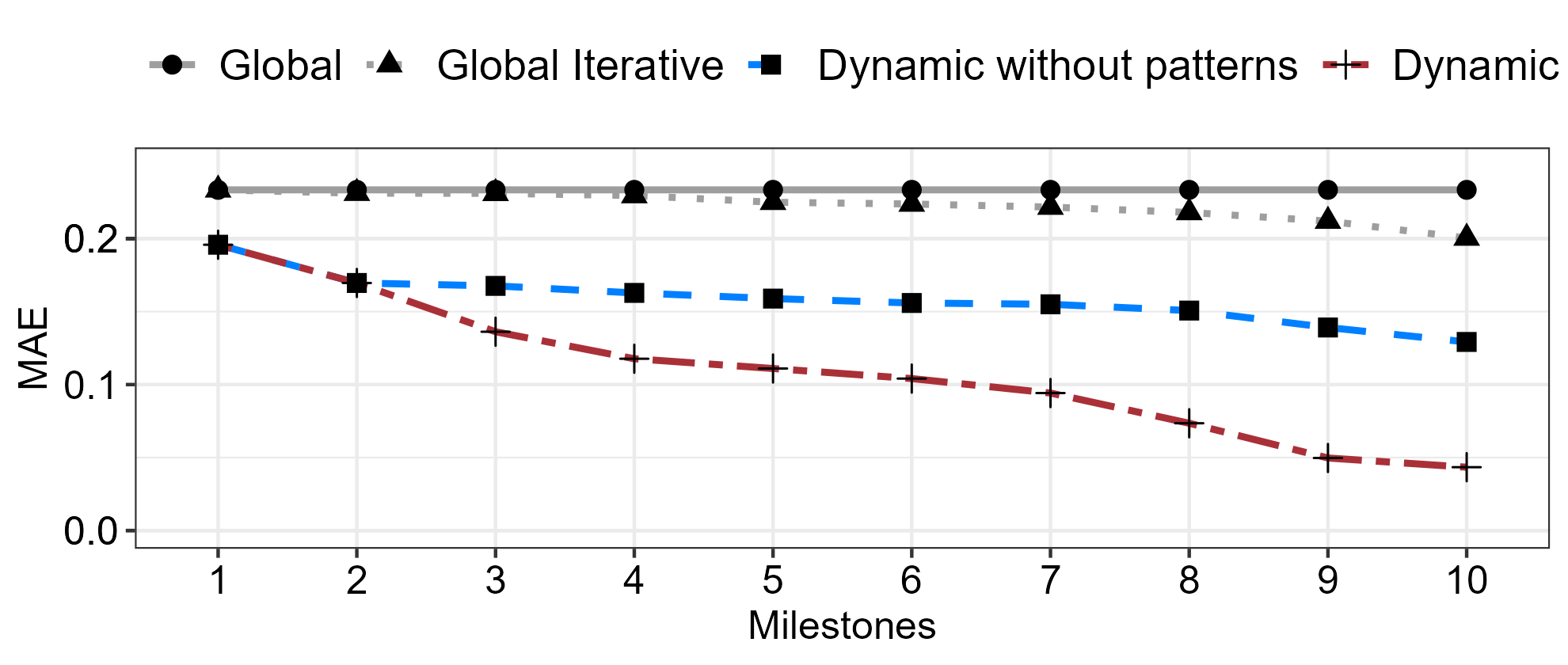}
            \caption{Mean Absolute Error over time}
            \label{fig:line-mae}
        \end{subfigure}
 \caption{Evaluation results obtained by the global, global iterative and dynamic Bayesian models over milestones (RQ1); dynamic with and without delay patterns (RQ2).}
        \label{fig:mode-comparison}
\end{figure*}

\subsection{Performance Measures}
We used the \emph{Mean Absolute Error} (MAE) and the \emph{Standardized Accuracy} (SA) as error measures; both have been recommended to compare the performance of effort estimation models~\cite{sarro2016multi, langdon2016exact}. MAE is defined as:
\setlength{\abovedisplayskip}{3pt}
\setlength{\belowdisplayskip}{4pt}
\begin{equation*}
MAE = \frac{1}{N} \sum_{i=1}^{N} | \ Actual \ BRE_{i} - Estimated \ BRE_{i} \ |
\end{equation*}

where $N$ is the number of epics used for evaluation, $Actual \ BRE_{i}$ is the actual delay measured in BRE, and $Estimated \ BRE_{i}$ is the predicted BRE value, for an epic $i$. SA is based on MAE and compares an effort estimation model against random guessing:
\vspace{-0.12mm}
\setlength{\abovedisplayskip}{4pt}
\setlength{\belowdisplayskip}{4pt}
\begin{equation*}
SA = \left( 1 - \frac{MAE}{MAE_{rg}} \right) \times 100 
\end{equation*}

where $MAE$ is defined as the MAE of the model that is being evaluated and $MAE_{rg}$ is the MAE of a large number of random guesses. SA represents how much better the model performs than random guessing. We used the unbiased exact calculation of $MAE_{rg}$ as proposed by Langdon et al.~\cite{langdon2016exact}. A lower $MAE$ and higher $SA$ imply better predictive performance.

To evaluate informativeness of the predictions of the Bayesian model (RQ4), we measured the relative width ($RWidth_{90}$) of the 90\% credible intervals~\cite{jorgensen2021evaluation}. A narrower interval (i.e. lower $RWidth_{90}$) is more informative.

To compare model performance, we tested the statistical significance of the evaluation results using the Wilcoxon Signed Rank Test~\cite{arcuri2014hitchhiker}. We applied the non-parametric Vargha and Delaney's $\hat{A}_{12}$ statistic~\cite{arcuri2014hitchhiker}, which is commonly used as effect size measure in effort estimation~\cite{sarro2016multi}.

\subsection{Results}

\textbf{RQ1: Benefits of dynamic prediction}

Figure~\ref{fig:mode-comparison} presents the evaluation results of the global, global iterative and dynamic modes of the Bayesian model for predicting the overall delay (in BRE) over milestones. Averaging across epics, the dynamic mode achieves 66--92\% SA and 0.19--0.04 MAE over milestones. Over time, the dynamic mode consistently outperforms the global mode by 12--57\% (SA) and 16--81\% (MAE), and the global iterative mode by 12--44\% (SA) and 16--78\% (MAE). The Wilcoxon test shows that the improvements achieved by the dynamic mode are significant ($p < 0.001$) with medium to large effect sizes ($\hat{A}_{12} = [0.65, 0.81]$). \emph{This indicates that the dynamic mode significantly improves global and global iterative modes right from the first milestone on.} \\

\textbf{RQ2: Benefits of delay patterns}

The dashed lines in Figure~\ref{fig:mode-comparison} present the evaluation results of the dynamic Bayesian model learned with and without delay patterns as input feature. At the first two milestones, the dynamic model learned using patterns provides the same estimations as the dynamic model learned without patterns. This is caused by the fact that the pattern clustering label becomes available from the third milestone on (i.e., when there is a series of two or more previous milestones to classify). Then, from the third milestone on, the dynamic model learned using patterns consistently improves the dynamic model without patterns by 9--20\% (SA) and 19--66\% (MAE). The improvements achieved by using delay patterns are significant with medium effect size ($\hat{A}_{12} = [0.64, 0.69]$). \emph{This indicates that the use of delay patterns leads to significant improvements in predictive performance, from the third milestone on.} \\

\textbf{RQ3: Comparison with SoTA baselines}

\begin{figure}[b!]
    \centering 
  \includegraphics[width=\linewidth]{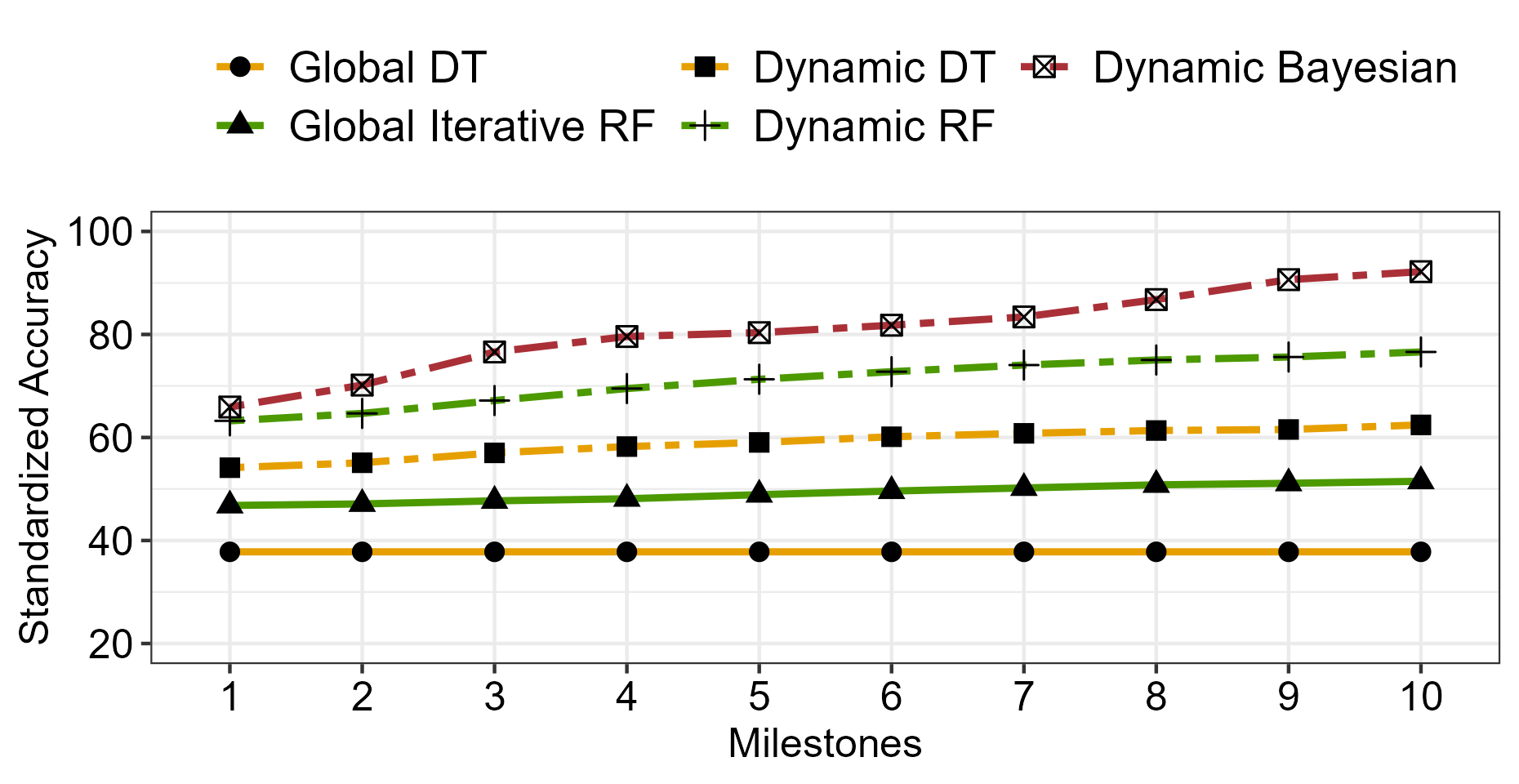}
  \caption{Comparison of our dynamic Bayesian model with SoTA baselines in global and dynamic modes (RQ3). `Global DT' and `Global Iterative RF' are the global Decision Tree~\cite{choetkiertikul2015threshold} and global iterative Random Forests~\cite{choetkiertikul2017predicting} learned using features from related work. `Dynamic DT' and `Dynamic RF' are the dynamic Decision Tree and dynamic Random Forests learned using our features from Table~\ref{tab:proxies}.}
  \label{fig:model-comparison}
\end{figure}

Figure~\ref{fig:model-comparison} presents the results of our dynamic Bayesian model compared to the SoTA baselines, represented by the Decision Tree~\cite{choetkiertikul2015threshold} and Random Forests~\cite{choetkiertikul2017predicting} models, in global and dynamic modes. The solid lines show the results of the Decision Tree and Random Forests models in their original, global mode using features from Choetkiertikul et al.~~\cite{choetkiertikul2015threshold, choetkiertikul2017predicting}. The dashed lines show the results of the models in dynamic mode using our features from Table~\ref{tab:proxies}.

The dynamic Bayesian model consistently outperforms the SoTA baselines in both global and dynamic modes. Over time, dynamic Bayesian improves the global Decision Tree by 74--144\% (SA) and 44--87\% (MAE), and the global iterative Random Forests by 56--71\% (SA) and 34--84\% (MAE). The Wilcoxon test shows that the improvements achieved by dynamic Bayesian over the global SoTA baselines are significant with large effect size ($\hat{A}_{12} > 0.84$). Dynamic Bayesian also outperforms the dynamic Decision Tree by 22-48\% (SA) and 26--80\% (MAE), and the dynamic Random Forests by 4--20\% (SA) and 7--68\% (MAE) over milestones. The improvements of dynamic Bayesian over the dynamic Decision Tree and Random Forests are significant with effect sizes greater than 0.58. \emph{This indicates that the dynamic Bayesian model achieves significant improvements over the SoTA baselines.} 



Overall, the models in dynamic mode substantially outperform their counterparts in global mode. This highlights the benefits of dynamic predictions across models. \emph{Bayesian achieves the highest predictive accuracy and the largest overall increase in performance compared to the SoTA baselines.} \\


\textbf{RQ4: Impact of prediction time}

\begin{figure}
    \centering 
  \includegraphics[width=\linewidth]{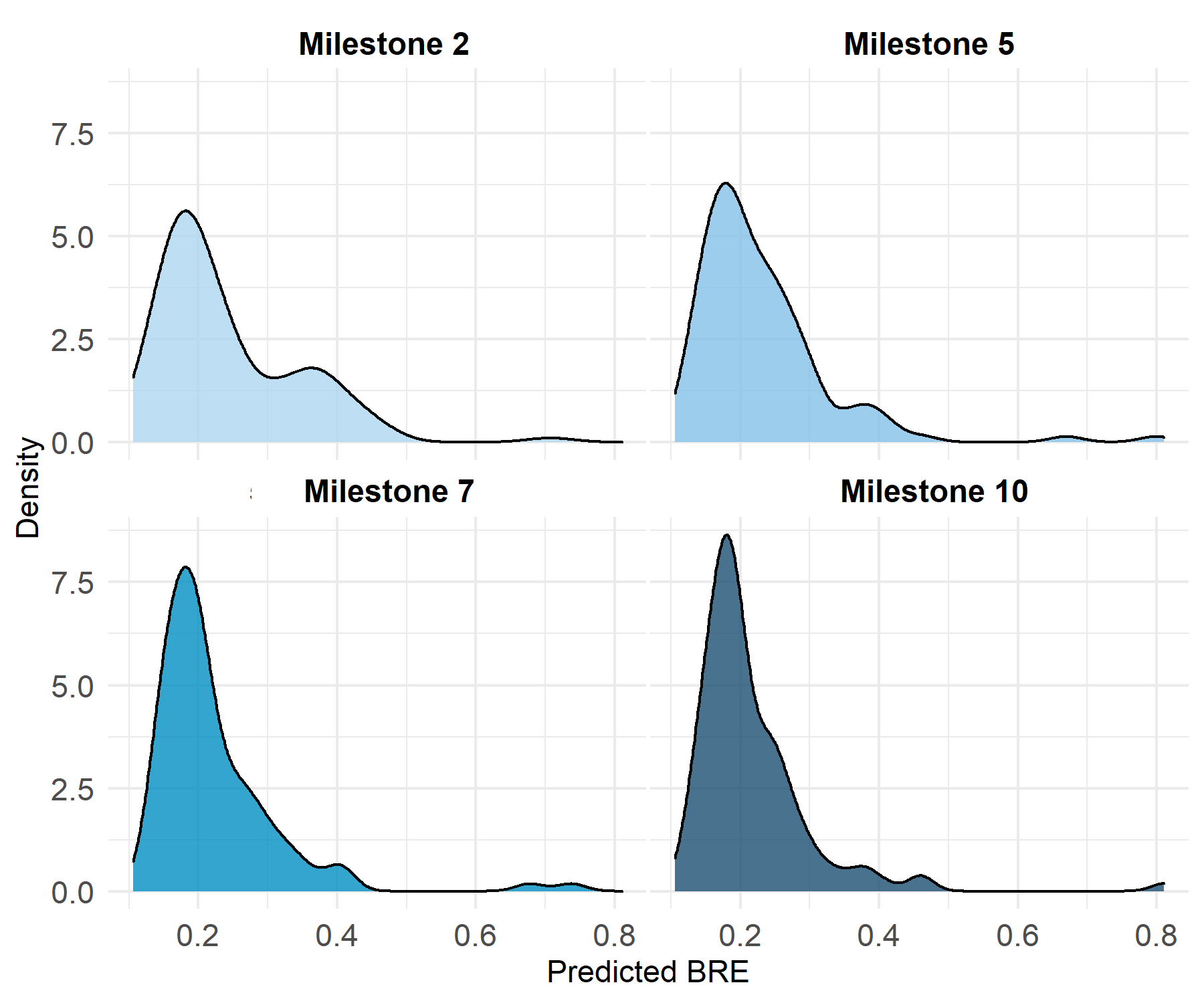}
  \caption{The estimated BRE distributions as updated by the dynamic Bayesian model across milestones (RQ4)}
  \label{fig:density}
\end{figure}

Figure~\ref{fig:density} shows how the estimated BRE distributions of the dynamic Bayesian model evolve over milestones 2, 5, 7 and 10. The prediction intervals become more narrow and sharp over time. The average $RWidth_{90}$ of the prediction intervals decreases from 1.14 at milestone 2 to 1.01 at milestone 5, 0.94 at milestone 7, and 0.89 at milestone 10. The Wilcoxon test shows that the changes in $RWidth_{90}$ over time are significant ($p < 0.001$) and the effect sizes are small to medium ($\hat{A}_{12} = [0.59, 0.68]$). \emph{This indicates that the dynamic Bayesian model is convergent, i.e. the predictions of the model become more certain and informative over time.}  

%% file: sections/discussion.tex
\section{Discussion}


\subsection{Main Findings}

\textbf{Delay patterns as input feature.} We found that the patterns identified at the case company are indicative of the overall project delay. This means that a similar pattern in historical data can provide an estimation for the future development of delay in an ongoing project. The patterns have shown their value in transport, and now in software development as well. They can be useful as input feature for delay prediction and rescheduling decisions. Our results demonstrate that the use of patterns leads to significant improvements of 9--20\% (SA) and 19--66\% (MAE) in the predictions of delay. The patterns in other organizations might differ from the four patterns identified at the case company. We expect that the number and shape of patterns will depend on the dataset being used. The patterns are essentially a reflection of recurring problems or abnormal behaviors that lead to delay in organizations. 


\textbf{Relationships with risk factors.} We characterized the patterns in terms of risk factors, as shown in Table~\ref{tab:rq1-2-characteristics}. Our statistical analysis reveals that the patterns show significant differences in various risk factors. Even though we cannot reason about causal links between the factors and patterns, the results of our factor analysis enable us to form hypotheses on the causes of delays. For example, the epics in Cluster 1 have a significantly higher number of outgoing dependencies, larger delivery scope and higher developer workload. We therefore hypothesize that large epics with many dependencies and overloaded developers are likely to exhibit a pattern similar to that of Cluster 1 and lead to major overall delay. Testing such hypotheses could lead to actionable insights and suggest delay mitigation measures. For a comprehensive view, we recommend the use of both epic- and story-level risk factors to characterize the patterns. Epic-level risks can provide high-level insights into problems related to the environment that the delivery takes place in. Story-level risks can give lower-level insights into problematic software tasks and collaboration challenges (e.g., user stories that have an abnormal waiting time are an indication of lack of team cooperation~\cite{choetkiertikul2015threshold}).

\textbf{Benefits of dynamic prediction.} Our results show that dynamic models significantly outperform their global and global iterative counterparts. The dynamic Bayesian model achieves improvements of at least 12--44\% (SA) and 16--78\% (MAE) right from the first milestone on. It also substantially outperforms the SoTA baselines. This highlights the benefits of dynamic prediction methods and indicates that existing, static methods are less suited to predict long-term delay. Existing models are not able to adequately incorporate changes occurring during project execution. Dynamic methods can effectively incorporate dynamic phenomena, resulting in increasingly more accurate and reliable schedule estimates over time. Dynamic prediction can therefore help teams detect risks throughout the project life cycle and react to delays in a more prudent fashion. This is especially valuable in development settings that are subject to constant change and where schedule overruns are a critical factor.

\textbf{Trade-off between prediction time and accuracy.} Our evaluation results show that the predictions of the dynamic model become more accurate and informative over time. We acknowledge that predicting at later times (at 70---100\% of the planned duration) may be less useful as it might be too late to change the outcome. However, the increased certainty may justify mitigation actions focused on handling a certain delay (e.g., postpone product launch, move features to other epic) instead of trying to catch up (by adding more resources). Furthermore, the dynamic approach achieves meaningful improvements right from the start of the project on. It improves the global and global iterative approaches by 12\% (SA) and 16\% (MAE) at 10\% duration, 19\% (SA) and 27\% (MAE) at 20\% duration and 29\% (SA) and 41\% (MAE) at 30\% duration. The improvements add up to 34\% (SA) and 37\% (MAE) at 50\% duration. The improvements obtained during the first half of the project are notable and can enable teams to take early measures against delay.

\textbf{Benefits of Bayesian methods.} In our comparison of the SoTA baselines in dynamic mode, Bayesian performs better than the Decision Tree and Random Forests models. Bayesian also achieves the largest overall increase in accuracy over time. This suggests that Bayesian is more effective in quantifying and updating the uncertainty of predictions over time. The results of RQ4 confirm this observation: the predictions of the Bayesian model become substantially more certain and informative over time. Unlike the other models, Bayesian provides detailed information about the uncertainty of an estimate in the form of a probability distribution. This can help organizations raise confidence in project plans.

\subsection{Future Work}

\textbf{Causal inference.} To improve the implementation of delay countermeasures, there is a need to better understand the causes of delays and delay patterns. An interesting direction for future research is to investigate why risk factors and delay patterns are related. This could be assessed by causal inference on individual patterns. Causal discovery (e.g., \cite{runge2019detecting}) could be used to learn a causal graph from the time series and identify the underlying causes of trends or fluctuations in the patterns. This can help software organizations to identify the causes of specific delays and estimate the effects of corrective actions beforehand. Another opportunity for future work is to map recurring peak moments in patterns onto development activities to identify key drivers of delay. Initial work in this direction has been carried out by Kerzazi and Khomh~\cite{kerzazi2014factors} and Kula et al.~\cite{kula2019releasing}. Both studies found that testing is one of the most time consuming activities and likely to result in delay.



\textbf{Systematic patterns.} The identified delay patterns might be affected by systematic effects that are calendar-related. Previous work (e.g., \cite{claes2018programmers, maddila2019predicting}) has shown the existence of such effects in software development work. An interesting opportunity for future research is to test for seasonality and model the time dependency of delay patterns using pattern matching. This would allow generalization over delay patterns and support the identification of systematic effects at different levels of time granularity. For instance, within-week dynamics due to day-of-the-week effects, and within-year dynamics affected by seasonal effects.


\textbf{Event-driven prediction.} Previous studies (e.g., \cite{kula2021factors, elbanna2015risks}) have found that software deliveries can be delayed by disruptive events, such as bugs and live incidents, that occur during project execution. Existing effort estimation models are static and therefore not able to incorporate such events into their predictions. Our dynamic model provides future research an opportunity to process incoming incidents as they occur. This would require updating of the model every time an incident or other notable event occurs. Previous studies~\cite{al2007dynarep, ameller2017towards} have recognized the potential of event-driven models for improving re-planning strategies in software projects.

\textbf{Delay propagation.} Currently, our dynamic model considers each software delivery independently and does not capture the interactions between dependent deliveries. However, a single delayed software delivery may cause a domino effect of secondary delays over dependent teams and projects. Future work should model the (dynamic) interrelation and propagation of delays across software deliveries. This could lead to more accurate estimates and a better understanding of the effects of delay propagation on delay patterns. Initial work in this direction has been carried out by Choetkiertikul et al.~\cite{choetkiertikul2015predicting}. They have shown that the use of networked data and collective classification leads to significant accuracy improvements. 

%% file: sections/threats-to-validity.tex
\section{Threats to Validity}


\textbf{Construct validity.} The data variables we consider may not capture the intended meaning of (concepts affecting) delay. This introduces possible threats to construct validity~\cite{ralph2018construct}. The delay measurements are derived from delivery dates and reported story points in the backlog management data. However, it might happen that teams do not take their delivery deadlines seriously and close their deliveries too early or too late. It is also possible that some teams do not follow the guidelines or principles for estimating story points. We tried to mitigate these threats by collecting real-world data from many epics and teams over a five year span. 

Another potential threat to our study is related to the milestone division of epics. We split the epics into regularly-spaced milestones based on completion rate. However, the milestones may not be a good match with the work pace of some teams. This might have led to a mixture of project phases within milestones and across epics, which would affect the results for the patterns in some deliveries. In practice, it would be more appropriate to split the epics based on iterations. 

\textbf{Internal validity.} 
The delay patterns that we condition our Bayesian model on may not reflect the situation in the test data. To mitigate this problem, we used time-based cross-validation to mimic a real prediction scenario. To compare models and verify our findings, we selected unbiased error measures and applied statistical tests~\cite{arcuri2014hitchhiker, menzies2006selecting}. 


\textbf{External validity.} 
External threats are concerned with our ability to generalize our results. We have analyzed 4,040 epics from 270 teams, which differ significantly in size, composition and product domains. However, we acknowledge that our data may not be representative of software projects in other organizations and open source settings. In other contexts, software deliveries might have a different setup following different collaboration practices. Replication of our work is needed to validate the findings in other settings and reach more general conclusions. 


%% file: sections/conclusion.tex
\section{Conclusions}
Modern agile software projects are volatile due to their iterative and team-oriented nature. Changes in risk factors and team performance trigger the need to re-assess overall delay risk throughout the project life cycle. Existing effort estimation models are static and not able to capture changes occurring during project execution. In this paper, we have proposed a dynamic effort estimation model for continuously predicting overall delay using delay patterns and Bayesian modeling. The model incorporates the context of the project phase and is finetuned based on changes in delivery performance over time. We apply our approach to real-world data from thousands of epics, identifying four intuitive delay patterns at \ING. The evaluation results demonstrate that:

\begin{enumerate}[]
    \item Delay patterns are indicative of the overall delay and useful as input feature for dynamic prediction.
    \item The dynamic model consistently outperforms global and global iterative approaches, and the SoTA baselines, even during early milestones (10--30\% of project duration).
    \item The predictions of the dynamic Bayesian model become substantially more certain and accurate over time.
\end{enumerate}

Overall, our results highlight the benefits of dynamic prediction methods that are able to learn from the time-dependent characteristics of software project delays. We identified several research areas calling for further attention, including causal inference, systematic effects, and delay propagation. Progress in these areas is crucial to better understand and manage delays in software projects. 



%% file: sections/data-availability.tex
\section{Data Availability}
The empirical data and source code used for this paper cannot be made publicly available due to an NDA. To encourage replication, we have described our model design step-by-step, and made our model summary and evaluation available in a replication package~\cite{supplemental}.